\documentclass[oribibl]{llncs}
\usepackage{graphicx}
\usepackage{url}

\newenvironment{keywords}{
       \list{}{\advance\topsep by0.35cm\relax\small
       \leftmargin=1cm
       \labelwidth=0.35cm
       \listparindent=0.35cm
       \itemindent\listparindent
       \rightmargin\leftmargin}\item[\hskip\labelsep
                                     \bfseries Keywords:]}
     {\endlist}

\title {Assessing Agile Transformation Success Factors}

\author {Amadeu Silveira Campanelli\inst{1} \and Florindo Silote Neto\inst{1} \and Fernando Silva Parreiras\inst{1}}
\institute{
    LAIS -- Laboratory for Advanced Information Systems, FUMEC University\\
    Av. Afonso Pena, 3880, Belo Horizonte 30130-009, Brazil \\
    \email{\{amadeu, florindo.silote\}@fumec.edu.br, fernando.parreiras@fumec.br} \\
}

\begin{document}

\maketitle
\begin{abstract}
				%    We like to think of writing an abstract as including five parts:
%    1) the context of the work,
Research on success factors involved in the agile transformation process is not conclusive and there is still need for guidelines to help in the transformation process considering the organizational context (culture, values, needs, reality and goals). 
%    2) the problem statement — specifically what the scope is,
The usage of success factors as a tool to help agile transformation raises the following research question: What are the success factors for an organization and their teams in preparation for the agile transformation process?
%    3) the methods — how the work was done,
This research presents an assessment of the organizational environment including the company's goals and the perception of the team members to provide awareness of
how the organization should prepare for the next steps in the agile transformation and a single case study for the assessment validation.
%    4) the results — what discoveries were made, and
The findings show that a company based in Chicago, USA, succeeded implementing customer involvement and self-organized teams but faces challenges with measurement models and training.
%    5) the conclusions — the significance and contributions of the work.
The main contributions of the research is the assessment of agile transformation success factors and the success factors difficulty ranking to be used by other organizations in their agile transformation processes.

\end{abstract}


\sloppy

\section[Introduction]{Introduction}
\label{sec:Introduction}

% It explains why there is a need or a problem and how the author will deal with it. The introduction usually motivates the present study, provides a literature review, and explains how the current work fits with what has gone before.

% P1 - What is the problem?
Organizations are searching for ways to overcome software development problems and achieve its business goals. Agile methods are an option many organizations have chosen to try to reach success~\cite{qumer2006crystallization,  jyothi2011effective, ayed2013quick, nishijima2013challenge, StateOfAgile2013}. As the number of organizations adopting agile methods continue to grow over the years~\cite{StateOfAgile2013}, the need to guide organizations through the adoption process is also growing~\cite{ayed2013quick, soundararajan_assessing_2013, javdani2015empirically}. Researchers noticed there is no unique agile adoption model and it is hard to adopt out of the box agile methods~\cite{ayed2013quick,  soundararajan_assessing_2013, javdani2015empirically, Campanelli2015, Gregory2016}. 
%TODO Needs a better connection to the challenges/success factors
The absence of systematic guidelines and frameworks for the agile adoption process considering the organization's context (culture, values, needs, reality and goals) directed  organizations to unsuccessful adoption paths and to failure~\cite{ayed2013quick, soundararajan_assessing_2013,gandomani2014human}. 

%Some researchers refer to these as adoption challenges or obstacles for agile adoption. We refer to them as success factors for agile adoption because [TBD]. They are important tools to understand to assess the level of [TBD] and the areas to invest during an adoption initiative. 

%Agile transformation/transition process (ATP) might require deep changes in the organization~\cite{gandomani2014human}.

% P2 - Why is it interesting and important?
During agile transition initiatives, organizations go through important transformations that impact its culture, hierarchy, management, environment and people~\cite{Gregory2016, NerurMM05}. Understanding the challenges or success factors for an agile transformation helps to prepare the people involved and increase chances of success.

% P3 - Why is it hard? (E.g., why do naive approaches fail?)
Agile transformation projects that do not consider the challenges to be faced along the way do not bring positive results to the organization and it can also create rejection from the professionals involved in it. Understanding the impacts of the changes in the organizational environment is an important step in the transformation process~\cite{NerurMM05}.

% P4 - Why hasn't it been solved before? (Or, what's wrong with previous proposed solutions? How does mine differ?)
Multiple articles in the literature list the success factors involved in an agile transformation process (ATP)~\cite{Gregory2016, ChowC08, TaylorGCMK08, gregory2015agile, gandomani2013obstacles} but there is still no direct guidelines of how to use these success factors in specific organizational contexts. Most of the articles found on success factor describe the success factors but do not specify ways to use those in the ATP. The usage of success factors as a tool to help in the agile transition initiatives raises the following research question: How to assess the current state of an organization and their teams in preparation for the agile transformation process?

% P5 - What are the key components of my approach and results? Also include any specific limitations.
We propose an assessment to provide awareness of the status of agile transformation success factors in the organization. Our approach starts with a consolidation of agile transformation success factors found in the literature and the mapping of success factors into phrases to be used in the assessment. The assessment was planned to evaluate both the organization leadership and team members points of view. Using these views of the organizational context, we applied  the Rasch algorithm~\cite{Lahrmann2011} to create a rank of difficulty of implementation of the success factors in the organization. We validated the assessment with a single case study in a software development company based in Chicago, USA.
The remainder of the paper is organized as follows. Section~\ref{sec:Assessment} introduces the success factor groups and proposes the agile transformation success factors assessment. The results of the assessment and the findings are discussed on Section~\ref{sec:Results}. Section~\ref{sec:Conclusions} outlines the conclusions and future work opportunities.

\section{Method}
\label{sec:Assessment}

%This research was conducted as a case study based on the guidelines of Runeson and H{\"o}st~\cite{Runeson09}. The stages of the research were the 

The agile transformation success factors assessment is a tool to provide awareness of how leadership evaluates the organization usage of the success factors and to help define goals for the ATP. It also considers the feedback from team members and the organizational context. The first step to create the assessment was to consolidate the success factors found in the literature into groups. Then, the success factors were mapped into representative phrases to be used in the assessment. 

The assessment was planned to evaluate the organization goals and the view of the people involved in the ATP of the current state of the organization. The organization's view of the success factors existence provides a reference for the target state, defining the level the company would like to achieve for each success factor. The team member's view of the success factors existence in the organization provides the current state of the organization. %Using the target and the current state we apply the Rasch algorithm~\cite{Lahrmann2011} to calculate the rank of difficulty of implementation of the success factors in the organization. 

%We propose an assessment to be used as a tool to provide awareness of how the organization evaluates itself, defines its goals and what should be the next steps for ATP considering the feedback from team members and the organizational context. Our approach starts with a consolidation of ATP success factors found in the literature and the mapping of success factors into phrases to be used in the assessment. The assessment was planned to evaluate the organization goals and view of the people involved in the transition process of the current state of the organization. The organization's view provides a reference of the target state, where the company should become. The team member's view provides the current state of the organization. Using the target and the current state we apply the Rasch algorithm~\cite{Lahrmann2011} to calculate the rank of difficulty of implementation of the success factors in the organization. We validate the assessment in a software development company based in Chicago, USA.

%\subsection{Agile Transformation Success Factors Groups}
%\label{sec:ATPGroups}

Success factor are the basis of the assessment since the evaluation occurs at their level. During the assessment definition, we found a need for a higher level of evaluation to allow a consolidated management view of the success factors.

We identified multiple success factors in the literature and aggregated them according to their concepts into terms and then into a set of groups. There were multiple references to the same concepts in different articles and we used terms we considered more representative of the concepts to aggregate all the references found. The six groups were proposed organizing the terms according to the areas of the organizations affected by them: customer, management, organization, process, team and tools. %The success factors and the groups used in this study are summarized in Table~\ref{tab:SuccessFactors1}. 
We select the Rash algorithm to be used to define how hard was to implement the sucess factors in the organization. The decision was based on the work of Lahrmann \textit{et al.}~\cite{Lahrmann2011} on maturity models using the Rasch algorithm where it was used to define which Business Intelligence capabilities are harder to be implemented and finally generate a maturity model clustering the capabilities into levels. In our case, we do not generate levels but instead we produce a ranking of the difficulty of implementing success factors in the organization.

\subsubsection{Case Study}

In order to validate the agile transformation success factors assessment we conducted a single case study in a software development company based in Chicago, USA called \textsf{Company A} in this research. The case study methodology was selected to allow an exploratory and qualitative view of the agile transformation success factors~\cite{runeson2009guidelines}. 

The case study was designed as a single case study to validate the assessment following guidelines by Runeson and H{\"o}st~\cite{runeson2009guidelines}. The case to be evaluated was the development process for \textsf{Company A} and the theory used in the case study was gathered from agile transformation success factors literature. Data collection was direct composed of interviews with the project sponsor and the major roles of the development process. 

\textsf{Company A} has 110 employees and it has been using agile methods and practices for about 9 years. Scrum is the agile method used as the base for the process. This organization works with a mature and consistent agile process and that is the reason we selected it to validate the assessment. We met with the agile adoption sponsor and director of the company to understand their ATP, how they currently work and also his impressions on the challenges they faced during agile transformation and their current challenges. He responded the assessment to define the target state from the organization point of view. 

We also interviewed 14 members of the organization with different profiles and skill sets: developers, designers, managers, testers and project managers/Scrum masters. Their responses to the assessment were considered the current state of the agile adoption in the organization. 71.4\% of the interviewees have been working with agile methods and practices for at least 6 years and 57.1\% of them have been working for \textsf{Company A} for at least 3 years. The teams represented by the interviewees were development (50.0\% of the interviewees), design (21.4\% of the interviewees), project management (14.3\% of the interviewees) and quality assurance (14.3\% of the interviewees).

Among the interviewees, 11 (78.6\%) considered the assessment very useful to provide the status of the success factors usage in the organization. Based in the expertise of the interviewees and the results collected in the interviews in \textsf{Company A}, we were able to validate the assessment.

%How can I assess the quality of my Results section? To make a self-assessment of your Results section, you can ask yourself the following questions.
%    * Have I expressed myself as clearly as possible, so that the contribution that my results give stands out for the referees and readers?
%    * Have I limited myself to only reporting the key result or trends that each figure and table conveys, rather than reiterating each value?
%    * Have I avoided drawing conclusions? (this is only true when the Results is an independent section)
%    * Have I chosen the best format to present my data (e.g. figure or table)? Have I ensured that this is no redundancy between the various figures and tables?
%    * Have I ensured that my tables of results are comprehensive in the sense that they do not exclusively include points that prove my point?
%    * Have I mentioned only what my readers specifically need to know and what I will subsequently refer to in the Discussion?
%    * Have I mentioned any parts of my methodology (e.g. selection and sampling procedures) that could have affected my results?
%    * Have I used tenses correctly? past simple for your findings (in the passive form), present simply (descriptions of established scientific fact)
\section{Results}
\label{sec:Results}

\begin{table}
	\centering
	\scriptsize
	\begin{tabular}{|r|l|p{5cm}|l|r|r|r|}
		\hline
		\textbf{Rank} & \textbf{Group} & \textbf{Success Factor} & \textbf{Logit} & \textbf{Error} & \textbf{Infit} & \textbf{Outfit} \\ \hline
		1 & Process & Measurement model  & 2.42 & 0.33 & 0.96 & 0.90 \\ \hline
		2 & Organization & Training  & 1.68 & 0.33 & 1.22 & 1.24 \\ \hline
		3 & Organization & Agile champions  & 1.11 & 0.35 & 0.48 & 0.58 \\ \hline
		4 & Organization & New mindset/roles   & 0.98 & 0.36 & 0.52 & 0.56 \\ \hline
		5 & Management & Changes in management style and decentralized decision making  & 0.85 & 0.37 & 0.19 & 0.23 \\ 
		\hline
		6 & Team & Distributed teams  & 0.85 & 0.37 & 1.78 & 1.04 \\ \hline
		7 & Organization & Knowledge sharing & 0.55 & 0.40 & 1.62 & 1.21 \\ \hline
		8 & Team & Technical activities/skills  & 0.55 & 0.40 & 1.71 & 1.20 \\ \hline
		9 & Organization & Business goals  & 0.21 & 0.43 & 0.33 & 0.36 \\ \hline
		10 & Process & Lightweight documentation  & 0.21 & 0.43 & 1.59 & 1.66 \\ \hline
		11 & Process & Process is compatible with the organizational context  & 0.21 & 0.43 & 0.48 & 0.54 \\ \hline
		12 & Team & Ability to build trustworthy relationships  & 0.21 & 0.43 & 0.72 & 0.74 \\ \hline
		13 & Team & Team involvement  & 0.02 & 0.45 & 0.72 & 0.66 \\ \hline
		14 & Organization & Incentives/motivation to adopt agile methods  & -0.20 & 0.48 & 0.62 & 0.59 \\ \hline
		15 & Organization & Communication flow in the organization &  -0.20 & 0.48 & 0.74 & 0.75 \\ \hline
		16 & Management & Management buy-in  & -0.44 & 0.50 & 0.38 & 0.32 \\ \hline
		17 & Organization & Coaching/mentoring  & -0.44 & 0.50 & 1.73 & 1.88 \\ \hline
		18 & Team & Collaboration  & -0.70 & 0.53 & 0.79 & 0.91 \\ \hline
		19 & Tools & Tool set  & -0.70 & 0.53 & 1.76 & 1.87 \\ \hline
		20 & Organization & Cultural changes  & -1.30 & 0.57 & 0.10 & 0.09 \\ \hline
		21 & Management & Changes in mind set of project managers &  -1.63 & 0.57 & 1.08 & 1.05 \\ \hline
		22 & Team & Self-organized teams  & -1.95 & 0.57 & 1.38 & 1.41 \\ \hline
		23 & Customer & Customer involvement &  -2.28 & 0.56 & 0.42 & 0.38 \\ \hline
	\end{tabular}
	\caption{Rasch algorithm results with success factors ordered from harder to easier to implement in \textsf{Company A}'s context.}	
	\label{tab:CaseStudyRaschResults}
\end{table}

We executed the Rasch algorithm using the Winsteps software~\cite{winsteps2016} based on the data collected in \textsf{Company A} to obtain the item calibration. The item calibration resulting of the Rasch algorithm classified the success factors in the assessment according to their difficulty to be implemented in \textsf{Company A}. 

Lahrmann \textit{et al.}~\cite{Lahrmann2011} proposed modifications to the data preparation for the Rasch algorithm for researches on maturity model for Information Systems (IS). We reproduced the same modifications in our research but at the end of the process we do not cluster the success factors in maturity models since that is not the goal of the research. 

The first modification was to use a 5-point Likert scale instead of the dichotomous scale (yes or no). The idea behind it is that in this case there is no right or wrong but instead, opinions to be expressed in a rating scale~\cite{Lahrmann2011}. The second modification was the usage of a representation of the desired or target state. In our research, the organization's goals for the success factor represents the desired or target state. The delta value representing the potential improvement is calculated based on the difference between current state and target state~\cite{Lahrmann2011}. Negative values would indicate that the current state is higher than the target state and that the success factor is already implemented in the \textsf{Company A} according to the defined goal.

The third modification proposed by Lahrmann \textit{et al.} was the coding of the delta value to encompass all negative values into a single category to simplify the expression of over-compliance~\cite{Lahrmann2011}. The coding used in this research is based on the work of Lahrmann \textit{et al.}~\cite{Lahrmann2011}.

In order to validate the results from the Rasch algorithm we used to the recommendations provided by Winsteps documentation~\cite{winsteps2016} and the guidelines used by Lahrmann \textit{et al.}~\cite{Lahrmann2011}. The fit statistics values (Infit and Outfit) are around 1.00 and satisfy the fit expectations validating the results. The results are summarized in Table~\ref{tab:CaseStudyRaschResults}.

Based on the results, the success factors that would be harder for \textsf{Company A} to implement are: measurement model, training, agile champions, new mindset/roles, and changes in management style and decentralized decision making. Furthermore, the results point out the there is already an adoption of customer involvement, self-organized teams, changes in mindset of project managers and cultural changes success factors by \textsf{Company A}.

\section{Conclusions}
\label{sec:Conclusions}

In this study, we consolidated success factors for agile transformation into groups and used them to propose the agile transformation success factors assessment. We validated the assessment in a case study in a company based in Chicago, USA. The agile transformation success factors assessment is a tool to help organizations to understand the current state of agile success factors in the organization based on their team members view and to prepare for the agile transformation process.

The assessment produces a success factors difficulty ranking according to the difficulty to implement the success factors in the organization considering its context (culture, reality, goals, hierarchy). For \textsf{Company A}, the ranking showed that measurement model, training, agile champions, new mindset/roles, and changes in management style and decentralized decision making are the harder success factors to be implemented. Meanwhile, customer involvement, self-organized teams, changes in mindset of project managers and cultural changes would be the easier to implement.

%The contributions of this research are: the success factors groups and the agile transformation success factors assessment. Researchers use the groups to understand how organizations are adopting agile methods. The assessment can be used by organizations to understand their current state regarding an ATP and plan on how to get to their target state. Future work will involve using the assessment in other organizations and also creating a way for the companies to assign importance/priority to the success factors according to the organizational context.

\bibliographystyle{splncs03}

\end{document}